\documentclass[fleqn,twoside]{article}
\usepackage{espcrc2}

\usepackage{epsfig}

\newcommand{\AmS}{{\protect\the\textfont2
        A\kern-.1667em\lower.5ex\hbox{M}\kern-.125emS}}

\title{Modeling Deep Inelastic Cross Sections in the Few GeV Region}

\author{A. Bodek\address[roc]{Department of Physics and Astronomy,
              University of Rochester,
              Rochester, New York 14618,  USA}
	and
	U. K. Yang \address[chic]{Enrico Fermi Institute, University
of Chicago,
              Chicago, Illinois 60637, USA}}

\begin{document}

\begin{abstract}
          We present preliminary results on
          simple modifications and corrections to GRV94 leading order
          parton distribution functions such that
          they can be used to model electron,muon and
          neutrino  deep-inelastic scattering cross sections at low energies.
          (Presented by Arie Bodek at NuInt01,
          Dec. 2001, KEK, Tsukuba, Japan).

\vspace{1pc}
\end{abstract}

\maketitle


Data from  atmospheric neutrino experiments~\cite{ATM} have
been interpreted  as evidence for $\nu_\mu \rightarrow \nu_{\tau}$ oscillations
with  $\sin^2 2\alpha > 0.88$  and
$1.6  \times 10^{-3} < \Delta m^2 < 4  \times 10^{-3}$
~${\rm eV^2}$.  These neutrino data are in the few GeV region.
Therefore, good modeling of $\nu_\mu$ and $\overline\nu_\mu$ cross sections
at low energies is needed.
The modeling of $\nu_\mu$ and $\overline\nu_\mu$ cross sections
is even more important for the more precise
next generation neutrino oscillations experiments.

These include MiniBooNE, MINOS, CNGS, and
experiments in the new neutrino facility to be constructed at the JHF
high intensity 50 GeV proton synchrontron in Japan.


The quark distributions in the proton
and neutron are parametrized as
Parton Distribution Functions (PDFs) obtained from
global fits to various sets of data at very high energies.
These fits are done within
the theory of Quantum Chromodynamics (QCD) in either leading order
(LO) or next to leading order (NLO). The most important data
come from deep-inelastic
e/$\mu$ scattering experiments on hydrogen and deuterium,
and $\nu_\mu$ and $\overline\nu_\mu$ experiments on nuclear targets.
In previous publications~\cite{highx,nnlo,yangthesis}
we have compared the predictions of the
NLO MRSR2 PDFs to deep-inelastic e/$\mu$ scattering data~\cite{slac}
on hydrogen
and deuterium from SLAC, BCMS and
NMC. In order to get agreement with the lower
energy SLAC data (down to $Q^{2} = 1$  GeV$/c^{2}$), and
at the highest values of $x$ ($x = 0.9$), we find
that the following modifications  to the MRSR2 PDFs must be
included.
\begin{enumerate}
       \item The relative normalizations between the various
             data sets and the BCDMS systematic error shift must
              be included~\cite{highx,nnlo}.
        \item Deuteron binding corrections need to be applied as
          discussed in ref.~\cite{highx}.
        \item The ratio of $d/u$ at high $x$ must be increased as
          discussed in ref.~\cite{highx}.
\item Kinematic higher-twist originating from target mass effects~\cite{gp}
are very large and
must be included.
\item Dynamical higher-twist corrections are smaller but also need
          to be included~\cite{highx,nnlo}.
\end{enumerate} In addition, our  analysis including QCD Next to NLO
(NNLO) terms
shows~\cite{nnlo} that most of the dynamical higher-twist corrections
that are needed to fit the data within a NLO QCD analysis  originate from
the missing NNLO higher order terms.
A recent calculation~\cite{Blum} also shows that dynamic higher twist
corrections
are very small.
If most of the higher-twist terms that are needed to
obtain agreement with the low energy data actually originate from target mass
effects and missing NNLO terms, then these
terms should be the same in  $\nu_\mu$  and e/$\mu$  scattering.
Therefore, low energy  $\nu_\mu$  data
should be described by the PDFs which are fit to high energy
data and are modified to include target mass and
higher-twist corrections to describe
low energy e/$\mu$ scattering data.
With this idea in mind, we find what modifications are need to be
applied to GRV94~\cite{grv94} leading order PDFs such that the PDFs
describe both high energy and low energy electron/muon data.

\begin{figure}[t]
\centerline{\psfig{figure=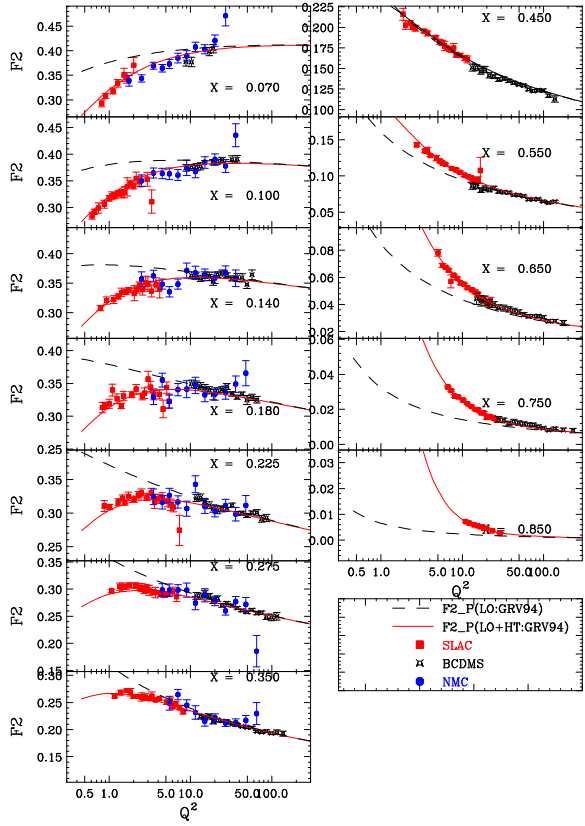,width=3.0in,height=4.3in}}
\caption{Electron and muon data on protons (SLAC, BCDMS and NMC) compared
to the predictions with GRV94
PDFs (LO, dashed line) and the modified GRV94 PDFs (LO+HT, solid line).}
\label{fig:f2p}
\end{figure}

In order to describe low energy data down to the photoproduction
limit ($Q^{2} = 0$), and account for both target mass and higher twist effects,
we find that the following modifications are required for the GRV94 LO PDFs.
\begin{enumerate}
\item  We increase the $d/u$ ratio at high $x$ as described in our previous
analysis~\cite{highx}.
The corrections to the $u$ and $d$ distributions
are described in detail in appendix A.
\item  Instead of
the scaling variable $x$ we use the
scaling variable $x_w = (Q^2+B)/(2M\nu+A)$ (or =$x(Q^2 +B)/(Q^2+Ax)$).
This modification was used in early fits to SLAC data~\cite{bodek}.
The parameter A provides for an approximate way to include target
mass effects and higher twist effects at high $x$,
and the parameter B allows the fit to be
used all the way down to the photoproduction limit ($Q^{2}$=0).
\begin{figure}[t]
\centerline{\psfig{figure=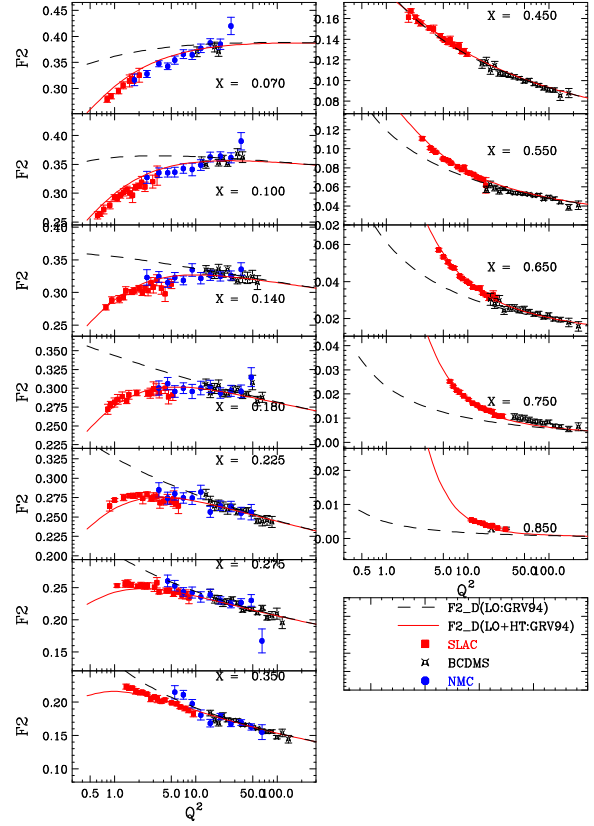,width=3.0in,height=4.3in}}
\caption{Electron and muon data on deuterons (SLAC, BCDMS and NMC) compared
to the predictions with GRV94
PDFs (LO, dashed line) and the modified GRV94 PDFs (LO+HT, solid line).}
\label{fig:f2d}
\end{figure}
\item  In addition as was done in earlier non-QCD based
fits~\cite{DL} to low energy data, we multiply all PDFs
by a factor $Q^{2}$ / ($Q^{2}$ +C). This is done in order for
the fits to describe both intermediate-$x$ data
and data in the photoproduction limit,
where the structure function
$F_{2}$ is related to the
photoproduction cross section according to
\begin{eqnarray}
     \sigma(\gamma p) = {4\pi^{2}\alpha_{\rm EM}\over {Q^{2}}}
          F_{2}
           = {0.112 mb~GeV^{2}\over {Q^{2}}}   F_{2} \nonumber
\nonumber
\end{eqnarray}
\begin{figure}[t]
\centerline{\psfig{figure=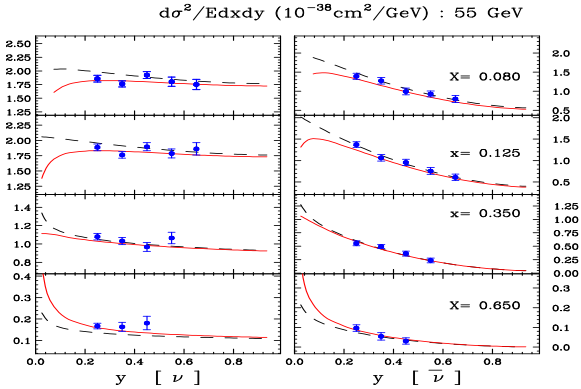,width=3.0in,height=2.0in}}
\caption{Comparison of representative CCFR $\nu_\mu$ and $\overline\nu_\mu$
charged-current differential cross sections~\cite{rccfr,yangthesis}
on iron at 55 GeV and the predictions of the GRV94 PDFs with (LO+HT,
solid) and without (LO, dashed)
       our modifications.}
\label{fig:ccfr}
\end{figure}
\item Finally,  we freeze the evolution of the GRV94 PDFs at a
value of $Q^{2}=0.24$ (for $Q^{2}<0.24$),
because the GRV94 PDFs are only valid down to $Q^{2}=0.23$ GeV/$c^{2}$.
\end{enumerate}

In our analysis, the measured structure functions
are corrected for the BCDMS systematic error shift and for
the relative normalizations between the SLAC, BCDMS
and NMC data~\cite{highx,nnlo}.
The deuterium data are also corrected
for nuclear binding effects~\cite{highx,nnlo}.

A simultaneous fit to both proton and deuteron data
yields A=1.735, B=0.624 and C=0.188.
Figures \ref{fig:f2p} and \ref{fig:f2d} show the SLAC, BCDMS
and NMC data as compared to the predictions of the standard leading GRV94 PDFs
(LO, dashed line) and with  our modifications (LO+HT, solid line).

Our value of C=0.188 is smaller than in other
analyses~\cite{DL}
to low $Q^{2}$ data. This is because we include both QCD evolution
and higher twist in our fits, while  QCD evolution is completely neglected
in these other  analyses~\cite{DL}.
We also compare the predictions with the standard  GRV94 PDFs
and with our modified GRV94 PDFs (LO+HT) to a few
representative high energy CCFR
$\nu_\mu$ and $\overline\nu_\mu$ charged-current
differential cross sections~\cite{rccfr,yangthesis} on iron
(these were not included in our fit).
In this comparison we use the
PDFs to obtain $F_{2}$ and $xF_{3}$ and correct for nuclear effects
on iron~\cite{rccfr,yangthesis}.
The parameterization for this nuclear effect is shown in appendix A
of this article.
The structure function $2xF_{1}$
is  obtained by using the $R_{world}$ fit from reference~\cite{slac},
as discussed in appendix A.
There is very good agreement with the neutrino data.

\begin{figure}[t]
\centerline{\psfig{figure=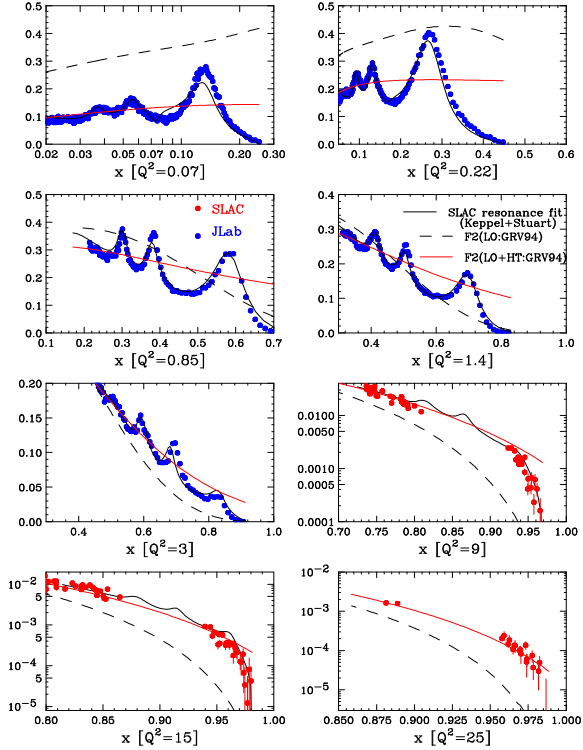,width=3.0in,height=3.9in}}
\caption{Comparison of a SLAC and JLab low energy
electron scattering
data in the resonance region (or fits to these data)
and the predictions of the GRV94 PDFs with (LO+HT, solid)
and without (LO, dashed) our modifications.}
\label{fig:resonance}
\end{figure}

According to Bloom-Gilman~\cite{bloom} duality and
if we use the $x_w$ scaling variable, the PDFs should
also provide a reasonable
description of the average value of the structure
functions in the resonance region. Figure \ref{fig:resonance} shows
a comparison between resonance data (from SLAC and Jefferson
Lab, or parametrizations of these data~\cite{jlab}) and
the predictions with the standard  GRV94 PDFs (LO)
and with our modified GRV94 PDFs (LO+HT).
There is good agreement with SLAC and JLab
resonance data down to $Q^{2}=0.07$ (although these data were not included in
our fit).  In the $Q^{2}=0$ limit, e.g. for $E_{CM}=2$ GeV,
the modified PDFs yield a photoproduction
cross section of 0.122 $\times$ 0.3/0.188 = 0.18 mb, which is
in good agreement with experimental data.

In order to have a full description of all charged current
$\nu_\mu$ and $\overline\nu_\mu$
processes,  the contribution from quasielastic
scattering~\cite{qe} must be added
separately at $x=1$.  One may chose to use these modified GRV94 PDFs
(LO+HT) to provide a description of all remaining inelastic
scattering processes,
including the resonance region, or one may chose to use them only
above a certain value of invariant mass $W$, and add the lower
mass resonances separately. Probably the best prescription is to
use these fits above the first resonance (e.g. above $W$=1.35 GeV) and
add the contributions  from
quasielastic and first resonance~\cite{rs} ($W$=1.23 GeV) separately.
In the resonance region at higher mass
       (e.g. $W$=1.7 GeV) there is a significant
contribution from the deep-inelastic continuum which is not
well modeled by the existing fits~\cite{rs} to
neutrino resonance data (and using
PDFs is better). Note that for
nuclear targets, nuclear corrections must also be applied.

In conclusion, we present the result of a first attempt at modifying
GRV94 PDFs (LO+HT) such that they provide a good description of e, $\mu$
and $\nu_\mu$ data both at low and high energies (including
the average of the cross section in the resonance
region). We have initiated a collaboration
with scientists at Jefferson Lab
to institute further improvements such as allowing for
different higher twist parameters for u, d, s, c
quarks (e.g. hydrogen
versus deuterium and valence versus sea). In addition,
one can multiply the PDFs by a modulating
function~\cite{bodek} A(W,$Q^{2}$)  to improve  modeling  in
the resonance region and comparing to Jefferson
lab data~\cite{jlab}. In addition, we plan to include
data on deuterium and heavier nuclear targets.
Note that because
of the effects of experimental resolution and Fermi motion (for
nuclear targets), a description of the average cross
section in the resonance region may be sufficient for some neutrino
experiments.
It is expected that there are some differences  between
the  form factors of resonances for $\nu_\mu$ and
$e/\mu$  scattering. As a test of our  approach, we will also
compare our predictions for $\nu_\mu$ scattering to
measured $\nu_\mu$ cross sections in the region of
first resonance~\cite{rs}, where the largest differences are
expected.

\appendix

\section{Appendix}

In leading order QCD (e.g.
GRV94 LO PDFs), $F_2$ for the scattering
of electrons and muons on proton (or neutron) targets is
given by the sum of quark
and anti-quark distributions (each weighted the
square of the quark charges):
\begin{eqnarray}
F_2(x) &=& \Sigma_i e_i^2 \left [xq_i(x)+x\overline{q}_i(x) \right].
\end{eqnarray}
Thus, our modified $F_2$ to describe low energy data down to
photoproduction limit is given by
\begin{eqnarray}
F_2(x)=\frac{Q^2}{Q^2+0.188} F_2(x_w),
\end{eqnarray}
where $x_w=x(Q^2+0.624)/(Q^2+1.735x)$.

In neutrino scattering we use the same modified scaling variable
and the same correction factor in $F_2$
and $xF_3$.

In the extraction of original
GRV94 LO PDFs, no separate longitudinal contribution was
included. The quark distributions
were directly fit to $F_2$ data. A full
modeling of electron and muon cross section requires
also a description of $2xF_1$.
We use a non-zero longitudinal $R$ in reconstructing $2xF_1$
by using a fit of $R$ to measured data.
Thus, $2xF_1$ is given by
\begin{eqnarray}
2xF_1(x) &=& F_2(x) (1+4Mx^2/Q^2) / (1+R).
\end{eqnarray}

In reality, a reconstruction of $2xF_1$ from
the values of $R$ and $F_2$ in neutrino scattering
is not as simple as in the case
of charged lepton scattering (because of charm
production). For charm production, the Bjorken scaling variable $x$ no longer
represents the fractional momentum carried by the struck quark
in the infinite momentum frame due to the non-zero heavy mass of the
charm quark
($m_c \sim 1.3$ GeV).  For charm
production processes, the variable $x$ is replaced
by the slow rescaling variable $\xi=(1+m_c^2/Q^2)x$.
Therefore, the structure functions for  the charm
production (cp) and non-charm production (ncp)
components are given by the following expressions.
\begin{eqnarray}
F_2^{cp+ncp}(x) & = & F_2^{ncp}(x)+ F_2^{cp}(\xi) \\
2xF_1^{cp+ncp}(x) & = & \frac{1+4M^2x^2/Q^2}{1+R^{ncp}(x)}
			F_2^{ncp}(x) \nonumber \\
                      &   &+ \frac{1+4M^2\xi ^2/Q^2}{1+R^{ncp}(\xi)}
			F_2^{ncp}(\xi)
\end{eqnarray}
In this model, the $R_{world}$ fit~\cite{slac} is used for $R^{ncp}$
and $R^{cp}$,
$R_{world}$ is parameterized by:
\begin{eqnarray}
R_{world}(x,Q^2) & = & \frac{0.0635}{log(Q^2/0.04)} \theta(x,Q^2) \nonumber \\
	         &   & + \frac{0.5747}{Q^2}-\frac{0.3534}{Q^4+0.09},
\end{eqnarray}
where $\theta = 1. + \frac{12 Q^2}{Q^2+1.0}
               \times \frac{0.125^2}{0.125^2 + x^2}$.

\begin{figure}[t]
\centerline{\psfig{figure=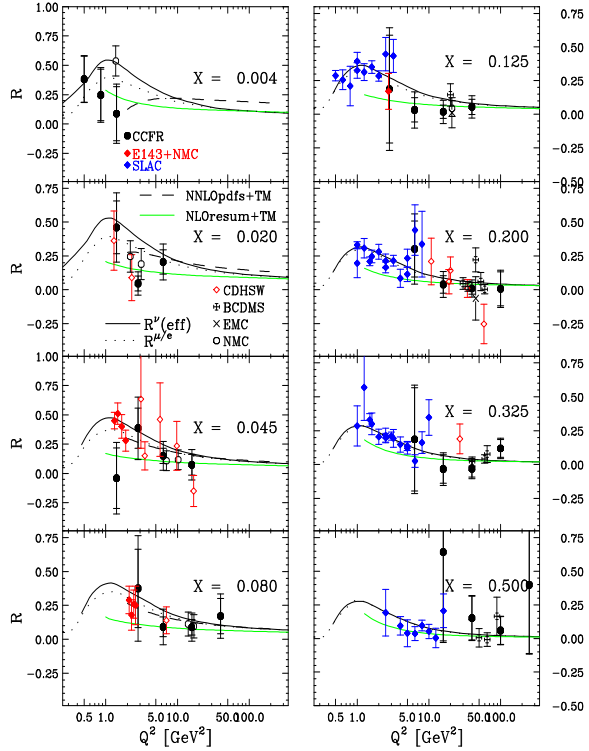,width=3.0in,height=3.9in}}
\caption{Comparison of $R$ data and $R_{world}$. The dotted line represents
$R_{world}$ (for electron/muon scattering). The solid line shows the 
$R_{eff}$ values for neutrino scattering,
reconstructed using Eq. 4 and 5.
The predictions with NLO resummation~\cite{resum}
and NNLO calculations~\cite{MRSNNLO} are also shown for comparison.}
\label{fig:rworld}
\end{figure}

The $R_{world}$ function provides a good
description of the world's data in the $Q^2>0.5$ and $x>0.05$ region
(where most of the $R$ data are available). Fig~\ref{fig:rworld}
shows a comparison of all available data on $R$ and $R_{world}$. 
$R_{world}$ is shown
as the dotted line.  Because of the effect
of the charm quark final state mass in charm production
in neutrino scattering,
 the effective $R$ for
neutrino scattering in the very low $x$ region (
$R_{eff}$ - solid line) is somewhat higher than $R_{world}$  (which
is a fit to
electron and muon scattering data). We use
Eq. 4 and 5 to construct $R_{eff}$ for neutrino
scattering.  However, this effect is not important at low neutrino
energies (which are below charm production threshold). We include
this effect because we want our formalism to be also correct at  high energies
(e.g. in the CCFR/NuTeV region).

Note that the $R_{world}$ function breaks down
below $Q^2=0.3$. Therefore, we freeze the function at $Q^2=0.35$ and introduce
the following function for $R$ in the $Q^2<0.35$ region.
The new function provides a smooth transition
from $Q^2=0.35$ down to $Q^2=0$ by forcing $R$ to approach zero at $Q^2=0$
as expected in the photoproduction limit (while
keeping a $1/Q^2$ behavior at large $Q^2$ and matching to $R_{world}$
at  $Q^2=0.35$).
\begin{eqnarray}
R(x,Q^2) & = & 3.207 \times \frac {Q^2}{Q^4+1} \nonumber \\
           &   & \times R_{world}(x,Q^2=0.35).
\label{eq:rmod}
\end{eqnarray}
In neutrino scattering the value of $R$ is required to approach zero
at $Q^2=0$ only for the vector
part of the interaction.
However, the overall contribution
from $R$ is expected to be small in this region. Therefore,
   it is reasonable to use Eq.~\ref{eq:rmod} for $R$ in
   both the electron/muon and neutrino scattering cases.

\begin{figure}[t]
\centerline{\psfig{figure=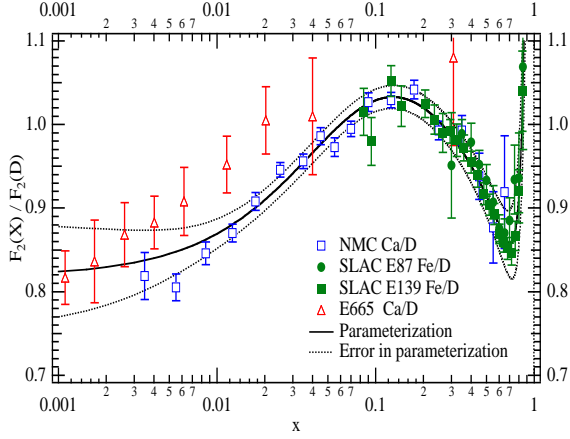,width=3.0in,height=2.3in}}
\caption{The ratio of $F_2$ data for heavy nuclear targets and
deuterium as measured in charged
lepton scattering experiments(SLAC,NMC, E665).
The band show the uncertainty of the parametrized curve from
the statistical and systematic errors
in the experimental data~\cite{selthesis}.}
\label{fig:nuclear_heavy}
\end{figure}

In the comparison with CCFR charged-current differential cross section
on iron, a nuclear correction for iron targets is applied.
We use the following parameterized function, $f(x)$
(fit to experimental electron and muon
scattering data for the ratio of iron to deuterium cross sections,
shown in Fig~\ref{fig:nuclear_heavy}),
to convert deuterium structure functions to (isoscalar) iron
structure functions~\cite{selthesis};
\begin{eqnarray}
f(x) & = & 1.096 -0.364x - 0.278e^{-21.94x} \nonumber \\
         &   & +2.772x^{14.417}
\end{eqnarray}

\begin{figure}[t]
\centerline{\psfig{figure=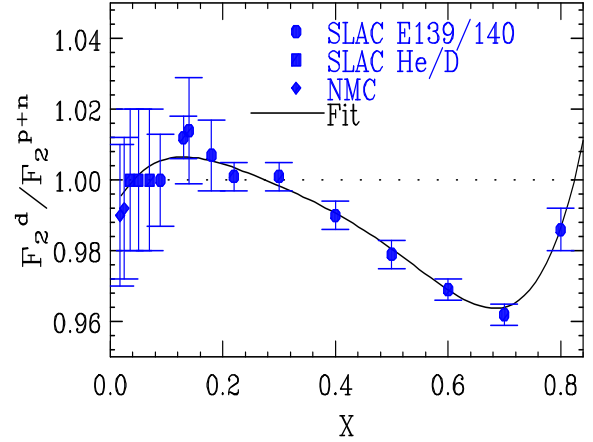,width=3.0in, height=2.3in}}
\caption{The total correction for nuclear
effects (binding and Fermi motion) in the deuteron,
    $F_2^d/F_2^{n+p}$, as a function of $x$, extracted from fits to
the nuclear dependence of SLAC $F_2$ electron scattering
data.}
\label{fig:f2dp}
\end{figure}

For the ratio of deuterium cross sections to cross
sections
on free nucleons we use the following function
obtained from a fit to SLAC data on the nuclear
dependence of electron scattering cross sections~\cite{yangthesis}.
\begin{eqnarray}
f(x) & = & (0.985 \pm 0.0013)\times (1+0.422x-2.745x^2 \nonumber \\
         &   & +7.570x^3  -10.335x^4+5.422x^5).
\label{eq:nucl-d}
\end{eqnarray}
This correction shown in Fig.~\ref{fig:f2dp} is only valid
in the $0.05<x<0.75$ region.  In neutrino scattering,
we use the same nuclear correction factor for $F_{2}$, $xF_{3}$
and $2xF_{1}$.

The $d/u$ correction for the GRV94 LO PDFs is obtained
from the NMC data for $F_2^D/F_2^P$.
Here, Eq.~\ref{eq:nucl-d} is used to remove nuclear binding effects
in the NMC deuterium $F_2$ data. The correction term, $\delta (d/u)$
is obtained
by keeping the total valence and sea quarks the same.
\begin{eqnarray}
\delta (d/u)(x)& = & -0.0161 + 0.0549x + 0.355x^2 \nonumber \\
                   &   & - 0.193x^3,
\end{eqnarray}
where the corrected $d/u$ ratio is $(d/u)'=(d/u)+\delta (d/u)$.
Thus, the modified $u$ and $d$ valence distributions are given by
\begin{eqnarray}
u_v' = \frac{u_v}{1+\delta (d/u) \frac{u_v}{u_v+d_v}} \\
d_v' = \frac{d_v+u_v \delta (d/u)}{1+\delta (d/u) \frac{u_v}{u_v+d_v}}.
\end{eqnarray}
The same formalism is applied to the modified $u$ and $d$ sea distributions.
Accidently, the modified $u$ and $d$ sea distributions (based on NMC data)
agree with the NUSEA data in the range of $x$ between 0.1 and 0.4.
Thus, we find that any futher correction on sea quarks is not necessary.

\end{document}